\documentclass[times]{cpeauth}

\usepackage{moreverb}

\usepackage{graphicx}
\usepackage{color}
\usepackage{listings}
\usepackage{bm}
\usepackage[hyphens]{url}
\usepackage{algorithm}
\usepackage{algorithmic}
\usepackage{amssymb}
\usepackage{todonotes}
\usepackage{booktabs}
\usepackage{amsmath}
\usepackage{multicol}
\usepackage{subfig}
\usepackage{epstopdf}

\usepackage{todonotes}

\newcommand{\dt}{{\Delta t}}

\title{Evaluation of DVFS techniques on modern HPC processors and accelerators 
for energy-aware applications}

\author{Enrico Calore\affil{1}\corrauth, 
        Alessandro Gabbana\affil{1},\\
        Sebastiano Fabio Schifano\affil{1}, 
        Raffaele Tripiccione\affil{1} \\
}

\address{
  \affilnum{1}University of Ferrara and INFN, Ferrara, Italy\break
}

\corraddr{
Via Saragat 1, I-44124 Ferrara, (Italy)
Email: enrico.calore@unife.it
Tel: +39 0532 974612
}

\begin{document}

\begin{abstract}
Energy efficiency is becoming increasingly important
for computing systems, in particular 
for large scale HPC facilities.
In this work we evaluate, from an user perspective, the use of Dynamic Voltage
and Frequency Scaling (DVFS) techniques, assisted by the power and energy
monitoring capabilities of  modern processors in order to tune applications for
energy efficiency.
We run selected kernels and a full HPC application on two high-end processors
widely used in the HPC context, namely an NVIDIA K80 GPU and an Intel Haswell CPU.
We evaluate the available trade-offs between energy-to-solution and 
time-to-solution, attempting a function-by-function frequency tuning.
We finally estimate the benefits obtainable running the full code on a HPC 
multi-GPU node, with respect to default clock frequency governors.
We instrument our code to accurately monitor power consumption and execution 
time without the need of any additional hardware, and we enable it to change 
CPUs and GPUs clock frequencies while running.
We analyze our results on the different architectures using a simple 
energy-performance model, and derive a number of energy saving strategies which
can be easily adopted on recent high-end HPC systems for generic
applications.

\end{abstract}



\maketitle


\section{Introduction}
\label{sec:intro}

The performances of current HPC systems are increasingly bounded by their energy 
consumption, and one expects this trend to continue in the foreseeable future.
On top of technical issues related to power supply and thermal 
dissipation, ownership costs for large computing facilities 
are increasingly dominated by the electricity bill.
Indeed computing centers are considering the option to charge not only 
running time, but also energy dissipation, in order to encourage users to 
optimize their applications for energy efficiency.

For HPC users there are different approaches to reduce the energy cost 
of a given application. 
One may adjust available power-related parameters of the systems on which 
applications run~\cite{freq-opt,freq-opt-gpu}, or one may decide to tune 
codes in order to improve their energy efficiency~\cite{code-energy-gpu}. 
Following the former approach, recent processors support \textit{Dynamic Voltage
and Frequency Scaling} (DVFS) techniques, making it possible to tune 
the processor clock frequency.
The software approach, on the other side, may be a winner in the long run, 
but it involves large programming efforts and also depends on reliable and 
user-friendly tools to monitor the energy costs of each kernel of a 
possibly large code.

This papers offers a contribution in the first direction, but also, 
while not discussing in details the software approach to energy optimizations, 
considers several reliable and user-friendly software tools and performance 
models which may eventually support the second direction too.

We focus on simple techniques to monitor the energy costs of key 
functions of large HPC programs, assessing the energy saving and the corresponding 
performance cost that one can expect -- for a given application 
code -- by a careful tuning of processor (i.e. CPU and GPU) clocks on a 
function-by-function basis.
In other words, we explore the energy-optimization space open in a situation in 
which -- due to code complexity, or for other reasons -- one would like to 
leave the running code unchanged, or -- at most --  annotate it.

\subsection{Related Works}

Processors power accounts for most of the power drained by 
computing systems~\cite{powerpack} and DVFS techniques were introduced to  
tailor a processor clock and its supply voltage. 
This approach has an immediate impact on the processor drained power, since 
-- to first approximation -- the power dissipation of gates is given by:

$$ P_{avg} = fCV^2 + P_{static}$$

\noindent
where $P_{avg}$ is the average total power, $f$ the working frequency, 
$C$ the capacitance of the transistor gates, $V$ the supply voltage, while 
$P_{static}$ is the static power, drained also while idle, accounting for 
example for leakage currents.

Power reduction does not automatically translate to energy
saving, since the average power drain of the system has to be integrated over 
the application execution time (or \textit{time-to-solution}, $T_S$) to obtain
the energy consumed by the application (or \textit{energy-to-solution}, $E_S$):

$$ E_s = T_s \times P_{avg} $$

\noindent
so an increase in $T_S$ may actually increase $E_S$, in spite of a lower 
average power drain $P_{avg}$.
This is a well known phenomenon motivating the convenience in some cases to run
at maximum clock frequency, no matter the power needed, in order to reach the
solution as soon as possible and then enter a low power idle state:
this is the so called ``Race-to-idle''.

Nowadays DVFS is routinely used to ensure that processors stay within an allowed
power budget, in order to run applications without over-heating the 
processor, and allowing short frequency bursts when possible.
The potential of DVFS to reduce \textit{energy-to-solution} for a complete 
application has been questioned with several arguments~\cite{dvfs-returns}, e.g.
since DVFS acts mainly on dynamic power, the trend of increasing ratio between 
static vs dynamic power drain reduces its effectiveness in lowering 
\textit{energy-to-solution}.
However, DVFS techniques have recently gained fresh
attention in various research works especially in the context  of large parallel
HPC contexts~\cite{future-dvfs}, where saving even a small  fraction of a very
large energy budget may have a significant impact.

Research work has focused in particular on communication phases, when processes
within large parallel MPI applications stop their execution, waiting for data 
exchanges.
These phases are indeed the best candidates to lower the CPU clock with minimal 
impact on overall performance~\cite{cpu-scaling-mpi,automatic-dvfs-parallel}.
However, from the point of view of performances, it is good practice to
overlap communication and computation whenever possible~\cite{hpcs15}.
Consequently, in several applications, in particular lattice based computations
(such as the one we adopt as a benchmark in this paper), communication phases
are often fully overlapped with computation, making this strategy less 
effective.

Other research works focused on the possibility to reduce the clock 
frequency in the case of an imbalance between the computational load of 
multi-task applications, allowing to reduce the clock frequency only for idle 
ranks~\cite{pmac-green-queue}.
But again, for the broad range of massively parallel computations, one tries to 
balance the computational load evenly on each of the computing threads or 
processes, reducing the impact of this opportunity.

Finally, the widespread adoption of accelerators in HPC systems means that the 
largest fraction of the power drained by computing systems is no more 
ascribable to CPUs. 
For instance,  with the advent of multi-GPU compute nodes, where up to 8
dual GPU boards are hosted on a single computing node, up to $\simeq 75\%$
of the power could be drained by GPUs, with CPUs accounting for
only   $5-10\%$ of the total energy budget~\footnote{Percentages referred to a
computing node of the COKA Cluster (\url{http://www.fe.infn.it/coka/}) installed
at the University of Ferrara and obtained from the declared maximum possible
power drain of the system and of the installed processors and accelerators.}. 
Consequently, as recent GPUs improve their support of DVFS, in many cases 
allowing for a fine grained frequency 
selection~\cite{freq-opt-gpu,power-perf-gpu,dvfs-k20}, various studies focused 
on the optimization space made available by tuning GPU clock frequencies.

\subsection{Contributions}
In a previous work~\cite{uchpc15} we analyzed the tradeoff between computing 
performance and energy efficiency for a typical HPC workload on selected 
low-power systems. We considered low-power \textit{Systems on a Chip} (SoCs)
since they are intended for mobile applications, so they provide several
advanced power monitoring and power saving features, but also  in view of a
prospective usage as building  blocks for future HPC
systems~\cite{hpc-building-blocks}. Given the increasing interest in controlling
and reducing energy dissipation, recent high-end processors have started to
support  advanced power-saving and power-monitoring technologies, similar to
those early adopted in low-power processors
~\cite{power-measurement,power-haswell}.

In this paper we perform a similar analysis on state-of-the-art HPC 
compute engines (hi-end processors and accelerators) addressing 
energy-performance tradeoffs for typical HPC workloads.
For this purpose, we compare different architectures from the energy efficiency
point of view, adopting some workload benchmarks that we had used earlier to 
compare CPUs and various accelerators from 
the sole point of view of performances~\cite{sbac-pad13,uchpc14}.
We use the same application benchmark as in~\cite{uchpc15}, a Lattice Boltzmann 
code, widely used in CFD and optimized for several 
architectures~\cite{cafgpu13,europar15,ccpe16}, this workload is well suited
for benchmarking purposes because its two most critical
functions are respectively strongly memory- and compute-bound.
Lattice Boltzmann implementations were recently used in other works,
e.g. in~\cite{chip-mpi-energy-opt} presenting a similar analysis 
for the Intel Sandy Bridge processor.

At variance with previous works such 
as~\cite{automatic-dvfs-parallel,pmac-green-queue}, we do not attempt to provide
an automatic tuning of processors frequencies nor an automatic tool able to
identify where to apply  frequency scaling.
As it is the case for various widely known ``computational challenges'' 
applications, we assume the users to already know the code functions where most 
of the computing time is spent and moreover to already know which are the 
compute- or memory-bound parts.
This knowledge is assumed to derive e.g. from a previous performance 
optimization, which could be complemented with an additional
optimization towards energy efficiency.
We aim therefore with this work to offer an in-depth overview of 
possible energy optimization steps available on recent common HPC architectures,
exploitable by HPC users that are already familiar with performance 
optimizations of their codes.

In this paper we follow an approach close to the one presented 
in~\cite{freq-opt} for an Intel CPU, exploring the optimization space for all 
possible CPU frequencies, but considering a more recent Intel Haswell 
architecture and without utilizing custom power measurement hardware. 
We also take into account GPU architectures, as recently done also 
in~\cite{freq-opt-gpu}, but using the more recent K80 architecture and a larger
set of frequencies. 

Our hardware testbed is a high-end HPC node of the COKA (COmputing on 
Kepler Architecture) Cluster, hosted at the University of Ferrara.
Each node in this cluster has 2 $\times$ Intel Haswell CPUs and 8 $\times$ 
dual GPU NVIDIA K80 boards.
We use two implementations of the same algorithm, previously optimized 
respectively for CPUs and GPUs~\cite{europar15,ppam15}, with different 
configurations and compilation options.

We make an extensive set of power and performance measurements of the critical 
kernels of the code, varying clock parameters for the hardware 
systems that we test. Using this information we assess the available 
optimization space in terms of energy and performance, and the best trade-off 
points between these conflicting requirements. 
We do so for individual critical routines, as we try to understand the observed 
behavior using some simple but useful combined models of performance, power and
energy.

We then investigate the possibility of tuning the clock frequency on a 
function by function basis, while running a full simulation, in order to run 
each function as energy-efficiently as possible.
We discuss our findings with respect to a simple energy-performance 
model. 
Finally, we demonstrate the actual energy saving potential, measuring the 
different energy consumptions of a whole computing node, while performing a full
simulation at different GPU clock frequencies.

\section{The application benchmark}
\label{sec:lbm}

Lattice Boltzmann methods (LB) are widely used in computational fluid dynamics,
to describe flows in two and three dimensions. LB methods 
-- discrete in position and momentum spaces -- are based on the synthetic 
dynamics of {\em populations} sitting at the sites of a discrete lattice. 
At each time step, populations hop
from lattice-site to lattice-site and then incoming populations {\em collide}
among one another, that is, they mix and their values change accordingly. 
LB models in $n$ dimensions with $p$ populations are labeled as $DnQp$; we
consider a $D2Q37$ model describing the
thermo-hydrodynamical evolution of a fluid in two dimensions, and
enforcing the equation of state of a perfect gas ($p = \rho T$)~\cite{JFM,POF};
this model has been used for large scale 
simulations of convective turbulence (see e.g.,~\cite{noi1,noi2}).
A set of populations ($f_l({\bm x},t)~l = 1 \cdots 37$), 
defined at the points of a discrete and regular lattice and each having a given 
lattice velocity $\bm c_{l}$, evolve in (discrete) time according to the 
following equation:
\begin{equation}
 f_{l}({\bm x}, t+\dt) = f_{l}({\bm x} - {\bm c}_{l} \dt,t) 
 -\frac{\dt}{\tau}\left(f_{l}({\bm x} - {\bm c}_{l} \dt,t) - f_l^{(eq)}\right)
\label{eq:master2}
\end{equation}
The macroscopic variables, density $\rho$, velocity $\bm u$  and temperature $T$
are defined in terms of the $f_l(x,t)$ and of the $\bm c_{l}$s;
the equilibrium distributions (${f}_{l}^{(eq)}$) are themselves a function of
these macroscopic quantities~\cite{sauro}. 
In words, populations drift
from different lattice sites ({\em propagation}), according to the value of
their velocities and, on arrival at point ${\bm x}$, they change their values
according to Eq.~\ref{eq:master2} ({\em collision}). One can show that, in
suitable limiting cases, the evolution of the macroscopic variables obeys the
thermo-hydrodynamical equations of motion of the fluid. Inspection of 
Eq.~\ref{eq:master2} shows that the algorithm offers a huge degree of 
easily identifiable parallelism; this makes LB algorithms 
popular HPC massively-parallel applications.

An LB simulation starts with an initial assignment of the populations, in
accordance with a given initial condition at $t = 0$ on some spatial domain, and
iterates Eq.~\ref{eq:master2} for each point in the domain and for as many
time-steps as needed.
At each iteration two critical kernel functions are executed:
%
i) {\sf propagate} moves populations across lattice sites 
       collecting at each site
       all populations that will interact at the next phase ({\sf collide}).
       Consequently, {\sf propagate} moves blocks of memory locations allocated
       at sparse addresses, corresponding to populations of neighbor cells;
%
ii) {\sf collide} performs all mathematical steps associated to 
      Eq.~\ref{eq:master2} in order to compute the population values at each
      lattice site at the next time step. Input data for {\sf collide}  are the
      populations just gathered by  {\sf propagate}.  {\sf collide} is
      the floating point intensive step of the code.
%
%

%
It is very helpful for our purposes that {\sf propagate}, 
involving a large number of sparse memory accesses, 
is strongly memory-bound;
{\sf collide}, on the other hand, is strongly compute-bound, and the
performance of the floating-point unit of the processor is here the ultimate 
bottleneck.

\section{Experimental setup}
\label{sec:setup}

All our tests were run on a single node of the COKA Cluster at the 
University of Ferrara. Each node has 2 $\times$ Intel Haswell E5-2630v3 CPUs
and 8 $\times$ dual GPU NVIDIA K80 Boards.

We use processor-specific implementations of the  LB algorithm of
Sec.~\ref{sec:lbm}, exploiting a large fraction of the available parallelism of
the targeted architectures.
On the GPU we run an optimized CUDA code, developed for large scale 
CFD simulations~\cite{sbac-pad13,uchpc14},  
while on the CPU we run an 
optimized C version~\cite{caf13,iccs13} 
using AVX2 \textit{intrinsics} 
exploiting the vector units of the processor 
and OpenMP for multi-threading within the cores.

For all codes and for each critical kernel we perform runs at varying values 
of the clock frequencies, and log power measurements on a fine-grained time 
scale.
For both of our target processors we read specific hardware registers/counters
able to provide energy or power readings.
This simple approach has been validated with hardware power meters by third 
parties studies for recent NVIDIA GPUs and Intel CPUs~\cite{power-measurement}, 
showing that it produces accurate results~\cite{power-haswell,papi}.
In this approach we cannot monitor the power drain of the motherboard and other 
ancillary hardware; recent studies have shown 
however that the power drain of those components is approximately constant in 
time and weakly correlated with code execution~\cite{power-measurement,power-kernel}. 
\textit{In-band} measurements may also
introduce overheads, which in turn may affect our results.
In this regard, we checked that performance degradation due to 
power measurements is negligible and measured the power dissipation of the
sole measurement code, verifying that it is less than +5\% of the
baseline power dissipation.

We developed a custom library to manage power/energy data acquisition from
hardware registers; it allows benchmarking codes to directly start and stop 
measurements, using architecture specific interfaces, such as the 
\textit{Running Average Power Limit} (RAPL) for the Intel CPU and the 
\textit{NVIDIA Management Library} (NVML) for the NVIDIA GPU. Our library also
lets benchmarking codes to place markers in the data stream in order to have
an accurate time correlation between the running  kernels and the acquired
power/energy values. For added portability of the instrumentation code, we
exploited the PAPI Library~\cite{papi} as a common API for energy/power 
readings for the different processors, partially hiding architectural  details.
The wrapper code, exploiting the PAPI library, is available for download as Free
Software~\cite{papi-reader}.

Our benchmark codes are also able to change the processor clock frequencies
from within the application. The code version for Intel CPUs uses the
\textit{acpi\_cpufreq} driver of the Linux kernel~\footnote{As of Linux Kernel 
3.9 the default driver for Intel Sandy Bridge and newer CPUs is 
\textit{intel\_pstate}, we disabled it in order to be able to use the 
\textit{acpi\_cpufreq} driver instead.}, able to set a specific frequency on 
each CPU core by calling the \textit{cpufreq\_set\_frequency()} function.
The \textit{Userspace} cpufreq governor has to be loaded in advance, in order to
disable dynamic frequency scaling and to be able to manually select a fixed 
CPU frequency.
As shown later, we also tested other clock governors in our benchmarks, such as
the \textit{Performance}, \textit{Powersave} and \textit{Ondemand} ones.
The NVIDIA GPUs code changes the GPU processor and memory frequency, using the 
NVML library and in particular the \textit{nvmlDeviceSetApplicationsClocks()} 
function.

\begin{figure}[ht!]
 \centering
 \subfloat[ Power drain while executing 1000 iterations of the {\sf propagate} 
  function, followed by 1000 iterations of the {\sf collide} function on one of 
  the two GPUs of an NVIDIA K80 accelerator at a requested GPU clock frequency 
  of 875MHz. As power exceeds the  designed TDP (150W), frequency is 
  automatically reduced to limit the power drain. The plot also shows the 
  processor temperature.] 
 {
    \includegraphics[width=0.8\textwidth]{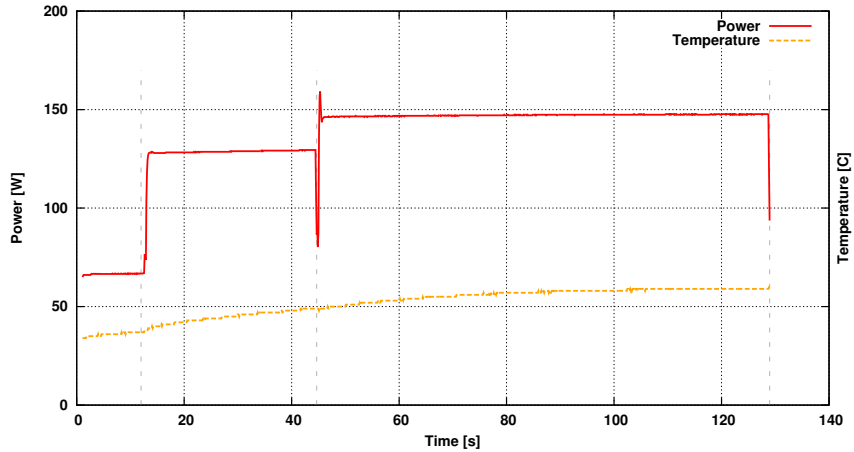}
    \label{fig:gpu-power}
 } \\
  \subfloat[Power drain on one of
  the two CPUs at a fixed CPU clock frequency of 2.4GHz for the same tests as in
  the previous panel; in this case we show separately the power drained by the 
  processor and by the memory system and their sum.] {
    \includegraphics[width=0.8\textwidth]{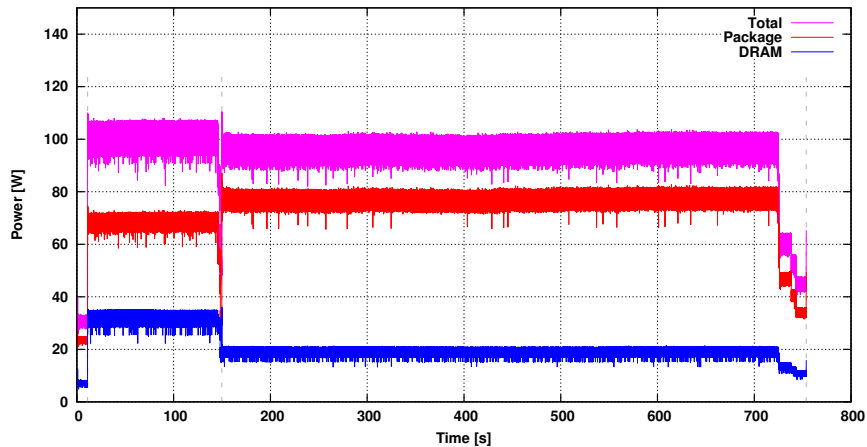}
    \label{fig:cpu-power}
 }
\caption{Plot of the power drain during the run of the two critical 
 kernels ({\sf propagate} and {\sf collide}) of our benchmark code; GPU data at 
 top, CPU data at bottom.}
 \label{fig:first-bench}
\end{figure}

The computationally critical section of a production run of our code performs 
subsequent iterations of: the {\sf propagate} function, the application of 
boundary conditions and then the {\sf collide} function as described in 
Sec.~\ref{sec:lbm}.
Boundary conditions depending on the details of the specific simulation,
have a marginal computational cost in typical simulations (of the order of 
$1-2\%$), so we neglect this computational phase in our detailed analysis (but 
the corresponding routines are included in the full code).

We have instrumented the two critical functions and
performed several test runs, monitoring the power/energy consumption.
All tests adopt double precision for floating
point data, the same lattice size ($1024 \times 8192$) and the same number of 
iterations ($1000$).

Fig.~\ref{fig:first-bench} shows
a sample of our typical raw data; 
Fig.~\ref{fig:gpu-power} shows the measured power drain
on one of the 2 GPUs of an NVIDIA K80 board at a fixed GPU frequency 
of 875MHz; 
additionally, GPU temperature during the run is obtainable (and plotted in 
the figure).
Fig.~\ref{fig:cpu-power} shows the same data for one Intel 
CPU at a fixed CPU frequency of 2.4GHz. On the Haswell
CPU we have two energy readings~\cite{power-haswell}: 
the \textit{Package Energy} (that we divide by the elapsed time to convert to 
power and plot in red) and the \textit{DRAM Energy} (again converted to power
and plotted in  blue).
The sum of the two readings --   the total power drained during the run -- is 
plotted in magenta.

The patterns shown in Fig.~\ref{fig:first-bench} are qualitatively easy to 
understand: we see a power surge during the execution of  
{\sf propagate} and a further large power figure during the execution of
{\sf collide}; Fig.~\ref{fig:gpu-power} tells us that on 
GPUs {\sf collide} is more power intensive than {\sf propagate} while 
Fig.~\ref{fig:cpu-power} informs us that in the latter routine the power drain 
of the memory system is comparatively larger, as expected in a memory-bound 
section of the code. 
As code execution ends, power drain quickly plummets.
An interesting feature in Fig.~\ref{fig:gpu-power} is the short power spike as 
{\sf collide} starts executing: in this case power briefly exceeds the TDP for 
this processor (150W) and a self-preservation mechanism reduces 
frequency to bring power down to acceptable levels; this 
mechanism will have an impact on our analysis, as discussed later. 

CPU data has larger fluctuations than GPU data, since samples for the former are
collected at 100Hz, while for the latter we have averages of longer time series,
collected at 10Hz, since for higher sample frequency, the NVML component of
the PAPI library starts to return repeated samples.
However, in both cases, time accuracy is enough to capture all interesting 
details.

We recorded data for all values of the clock
frequency of the processor at a fixed memory frequency of 2133 MHz (CPU) and
2505 MHz (GPU): indeed -- at variance with low-power processors~\cite{uchpc15}
-- the systems that we test run at a fixed frequency of the memory interface.
This data is the starting point for the analysis shown in the 
following~\footnote{The NVIDIA K80 memory clock frequency can also be set to 
324 MHz, but such a low frequency is designed to be useful only to reduce power
drain when the processor idles.}.

\section{Performance and Energy models}
\label{sec:models}

In this section, we consider figures-of-merit useful to assess the 
energy cost of a computation and to develop some models that will guide us 
in the analysis of our experimental data. 

We consider the {\em time-to-solution} ($T_S$) and {\em energy-to-solution} 
($E_S$, in Joule, defined as the average power $P_{avg}$ multiplied by $T_S$) 
metrics and their correlations as relevant and interesting parameters when 
looking for tradeoffs between conflicting energy and performance 
targets.
Other quantities -- e.g. the {\em energy-delay product} (EDP) -- have been 
proposed in the literature, in an attempt to define a single 
figure-of-merit;
however we think that correlations between several parameters 
better highlight the underlying mechanisms; other authors share this
attitude (see for instance~\cite{chip-mpi-energy-opt}).

We adopt a simple model that links $E_S$ and $T_S$, combining 
the \textit{Roofline Model}~\cite{roofline}, with a
simple power model for a generic processor. 
The \textit{Roofline Model} identifies in any processor based on the Von 
Neumann architecture two different subsystems, working together to perform a 
given computation: a compute subsystem with a given computational performance
$C$ in \textit{operations-per-second} or \textit{FLOPS/s} (if the workload is 
mostly floating-point)  and a memory subsystem, 
providing a bandwidth $B$ in \textit{words-} or
\textit{bytes-per-seconds}, between processor and memory.
The ratio $M_b = C/B$, specific of each hardware architecture, is 
the \textit{machine-balance}~\cite{machine-balance}. 
Looking now at a generic target application, the \textit{Roofline Model} 
considers that every computational task performs a certain number of 
operations $O$, operating on $D$ data 
items to be fetched/written from/onto memory. For any software function, the
corresponding ratio $I = O/D$ is the \textit{arithmetic intensity}, 
\textit{computational intensity} or \textit{operational 
intensity}~\cite{roofline}.

The \textit{Roofline Model} derives performance estimations combining 
\textit{arithmetic intensity} and \textit{machine-balance}. Using a 
notation slightly different from the one of the original paper~\cite{roofline}, 
one obtains that to first approximation --  e.g., neglecting scheduling 
details --
\begin{equation}
T_S \approx max[T_c, T_m] = max[\frac{O}{C}, \frac{D}{B}] = \frac{O}{C} max[1, \frac{DC}{OB}] =
                      \frac{O}{C} max[1, \frac{M_b}{I}].
\label{eq:maxT}   
\end{equation}
Eq. \ref{eq:maxT} tells us that $T_S$ is a decreasing functions of $C$, as 
naively expected, but only as long as $M_b/I \le 1$;
if this condition is not fulfilled, $T_S$ stays constant (and we say that the
task is memory-bound). To very good approximation, $C$ is proportional to the 
processor frequency $f$, so, varying $f$ also modulates $M_b$ and 
one derives that
\begin{equation}
T_S(f) \propto \frac{1}{f} max[1, \alpha f]
\label{eq:T}
\end{equation} 
with an appropriate constant $\alpha$. 

\begin{figure}[t!]
 \centering
\includegraphics[width=0.8\textwidth]{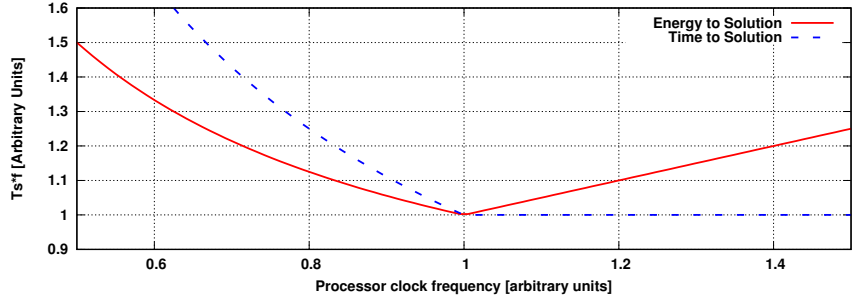}
\caption{Sketchy behavior of $E_S$ and $T_S$ as predicted by Eqs. \ref{eq:ET}
and \ref{eq:T}. All quantities in arbitrary units; the frequency value at
which $I = M_b$ is normalized to 1. }
\label{fig:sketch}
\end{figure}
To relate power dissipation $P$ and $f$ we assume a simple linear 
behavior, ignoring more complex effects such as the leakage current increase 
with temperature and the voltage change often associated with frequency scaling:
\begin{equation}
P(f) \approx m~f + P_s
\label{eq:P}
\end{equation}
where $P_s$ accounts for leakage currents and for the static power
drained by parts of the system not used in the specific task. 

\noindent
Combining Eq.~\ref{eq:T}  and Eq.~\ref{eq:P}, we have:
\begin{equation}
E_S = P(f) \times T_S(f) \propto (m+\frac{P_s}{f})~max[1, \alpha~f]
\label{eq:ET}
\end{equation}
This equation shows that, as we increase $f$, $T_S$ and $E_S$ both 
decrease provided that $\alpha f \le 1$; however, as soon as $\alpha f > 1$, 
increasing $f$ becomes useless, as $E_S$ starts to increase again while $T_S$ 
remains constant. 

This behavior, sketchly shown in Fig. \ref{fig:sketch}, tells us that the 
matching of $M_b$ and $I$, that in the \textit{Roofline Model} is relevant for 
performance, is an equally critical parameter for energy 
efficiency, as noticed also in~\cite{roofline-energy} and 
in~\cite{roofline-energy-gpu}: indeed, $M_b/I \approx 1$ 
gives the best performance and {\em at the same time} the lowest energy 
dissipation.

Seen from a different perspective we have a duality between complementary 
approaches: software optimizations for a specific architecture
change $I$ to adapt to a given $M_b$ while, changing clock 
frequencies we try to match $M_b$ to a given $I$.

\section{Data analysis}
\label{sec:dataAnalysis}

\begin{figure}[ht!]
 \centering
 \subfloat[\sf Propagate]{
   \includegraphics[width=0.8\textwidth]{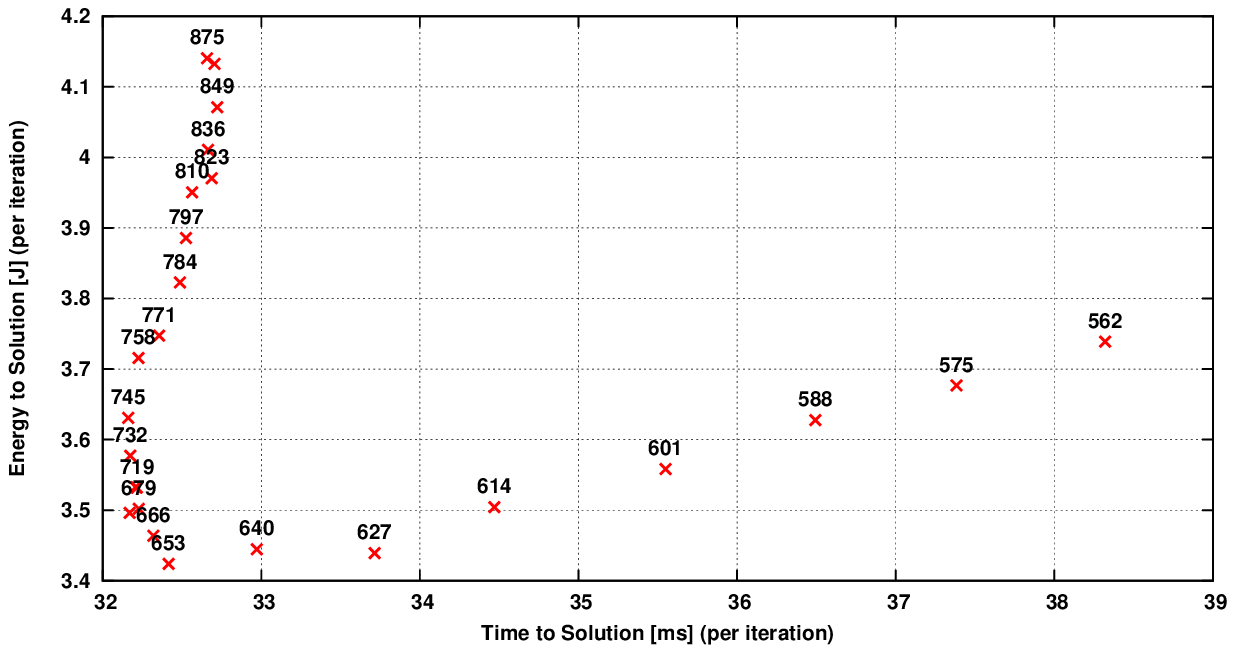}
   \label{fig:gpu-prop-et}
 }\\
 \subfloat[\sf Collide]{
   \includegraphics[width=0.8\textwidth]{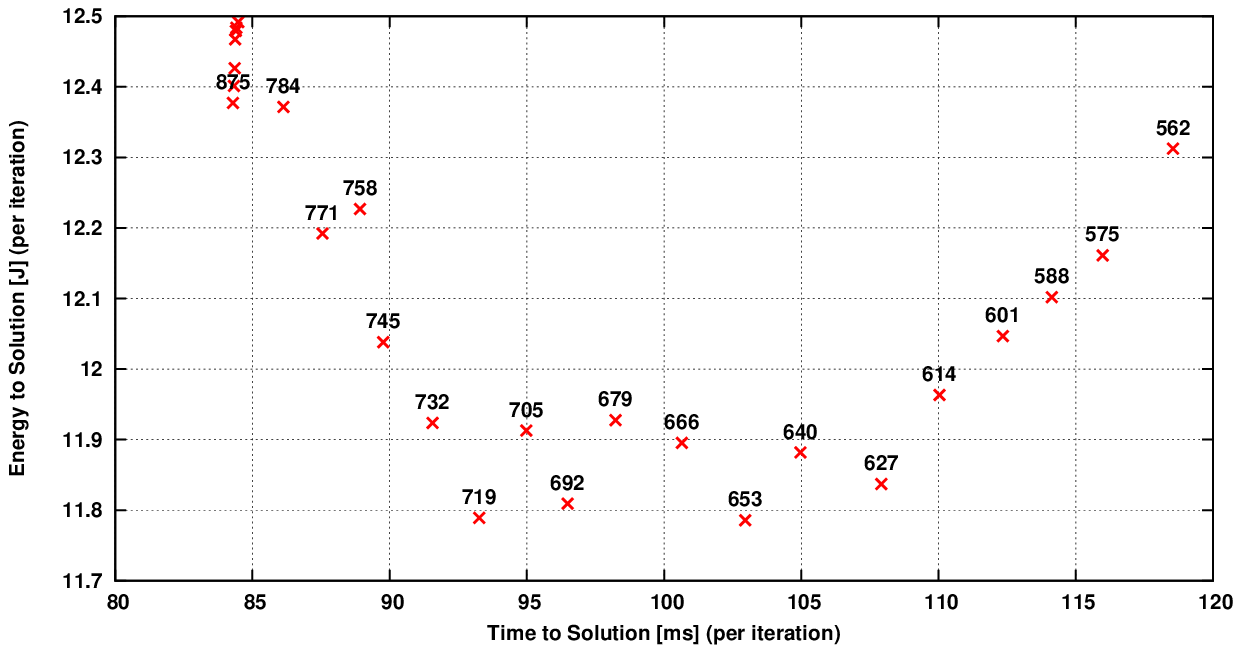}
   \label{fig:gpu-coll-et}
 }
 \caption{$E_S$ vs $T_S$ for the {\sf propagate} and {\sf collide} functions, 
 measured on the GPU; labels are the corresponding clock frequencies $f$ in 
 MHz.}
  \label{fig:gpu-et}
\end{figure}

\begin{figure}[ht!]
 \centering
 \subfloat[\sf Propagate]{
\includegraphics[width=0.8\textwidth]{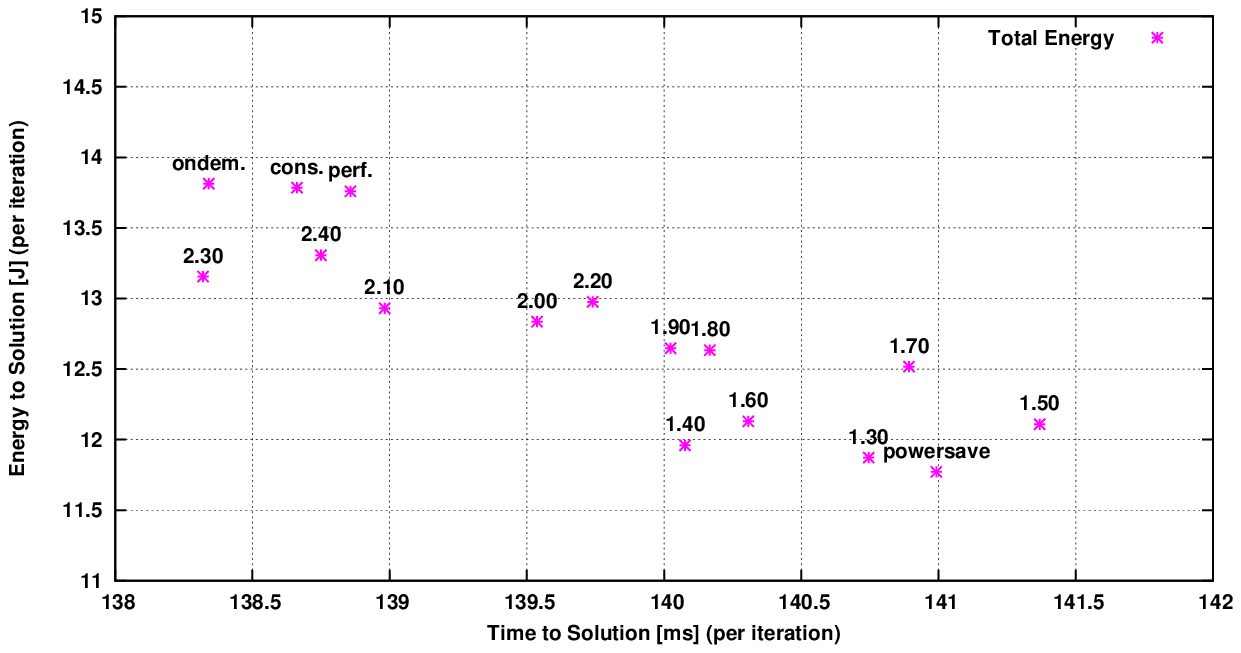}
\label{fig:cpu-prop-et}
 }
 
 \subfloat[\sf Collide]{
\includegraphics[width=0.8\textwidth]{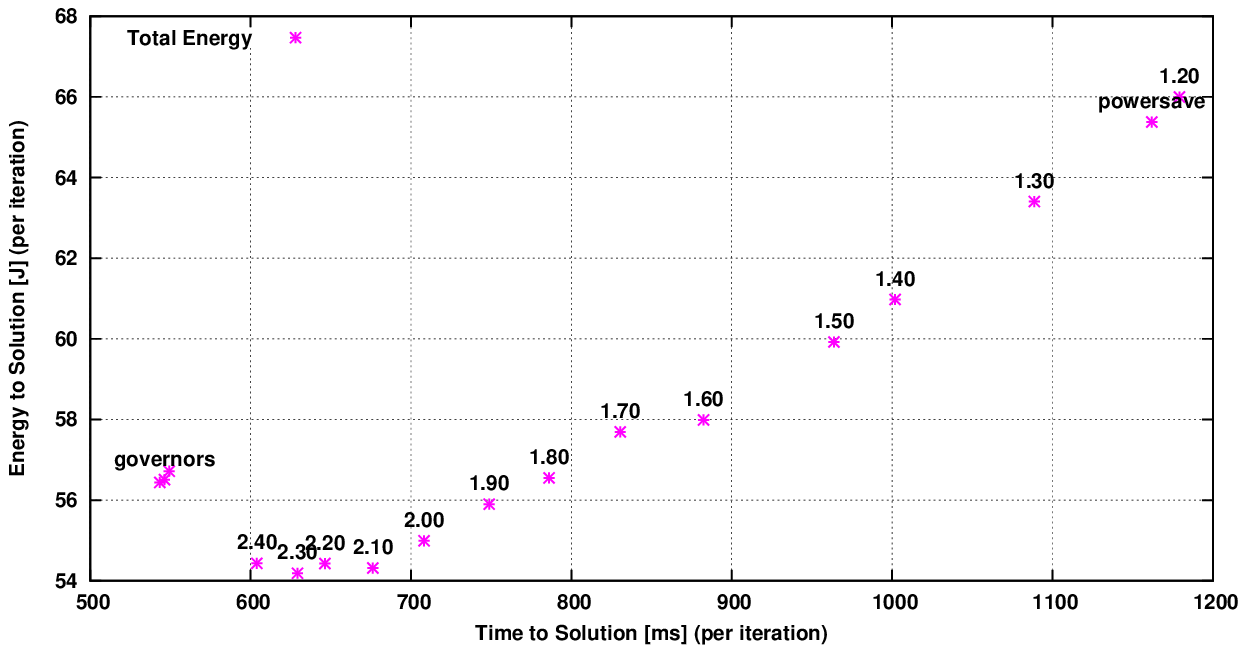}
 \label{fig:cpu-prop-et-zoom}
}
\caption{$E_S$ vs $T_S$ for the {\sf propagate} and {\sf collide} functions, 
measured on the CPU; labels are the corresponding clock frequencies $f$ in GHz.}
\label{fig:cpu-et}
\end{figure}

We plot our results for $E_S$ vs. $T_S$ in Figs.~\ref{fig:gpu-et} 
and~\ref{fig:cpu-et}, showing experimental values for the {\sf propagate} and 
{\sf collide} 
kernels, for both processors and for all available clock frequencies.

At the qualitative level, the results of Figs.~\ref{fig:gpu-et} 
and~\ref{fig:cpu-et} have a slightly diverging behavior across the two 
architectures; GPU data show a correlated decrease of $E_S$ and 
$T_S$ till, at some $E_S$, the trend reverses, with $E_S$ sharply
rising again. 

CPU data for {\sf propagate} on the other hand shows that $T_S$ is almost constant 
($T_S$ changes by $\approx~4\%$, while the clock
frequency changes by a factor 2), while $E_S$ increases by $\simeq 10\%$ as 
$T_S$ decreases. 
On the same processor, {\sf collide} shows a gentle correlated decrease of $E_S$
and $T_S$ for all frequencies that we are able to control. The best $T_S$ values 
shown in the plot are obtained using default cpufreq governors: 
\textit{Performance}, \textit{Ondemand} and \textit{Conservative}.
These default governors allows the processor to use so-called \textit{Turbo Boost} 
frequencies, enabling the processor to run above its nominal 
operating frequency, but this happens at a significant energy cost.

\begin{figure}[ht!]
 \centering
\includegraphics[width=0.8\textwidth]{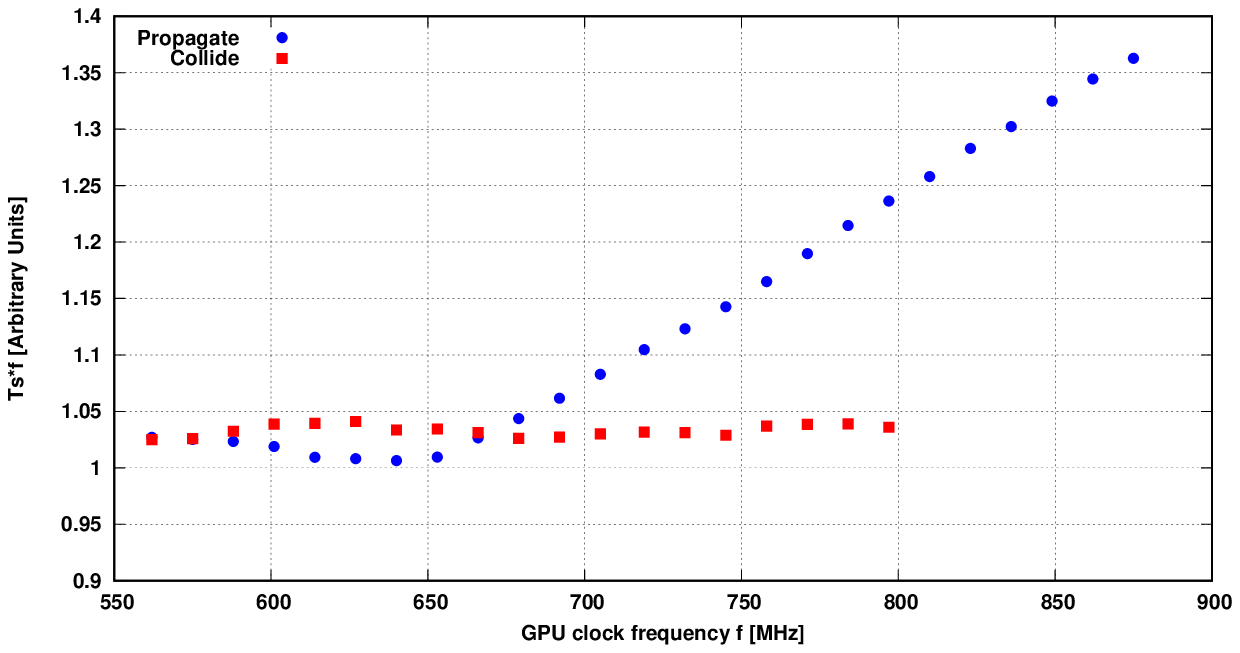}
\caption{$(f~T_S)$ as a function of $f$ for {\sf propagate} and
{\sf collide}, running on a GPU. $(f~T_S)$ is constant as long as 
$M_b/I \le 1$, corresponding to a compute-bound regime; increasing values of 
$(f~T_S)$ indicate a memory-bound regime.} 
\label{fig:timeVsFreq}
\end{figure}

We now compare our data with the model of the previous 
section, starting from an analysis of GPU data. A qualitative comparison of
data with the model suggests that both routines become memory-bound at
some value of the clock frequency; we check this assumption in Fig. 
 \ref{fig:timeVsFreq} where we show (in arbitrarily rescaled units) 
the product 
$(f~T_S)$ as a function of $f$ for both kernels: as long as $T_S$  scales as 
$1/f$, $(f~T_S)$ has to take a constant value.  We see that for {\sf collide}
this is true up to $f < 800$ MHz while for {\sf propagate} the change happens
already at $f \simeq 650$ MHz. Our model would then imply that at those
frequencies the two functions become respectively memory-bound.

\begin{figure}[ht!]
 \centering
\includegraphics[width=0.8\textwidth]{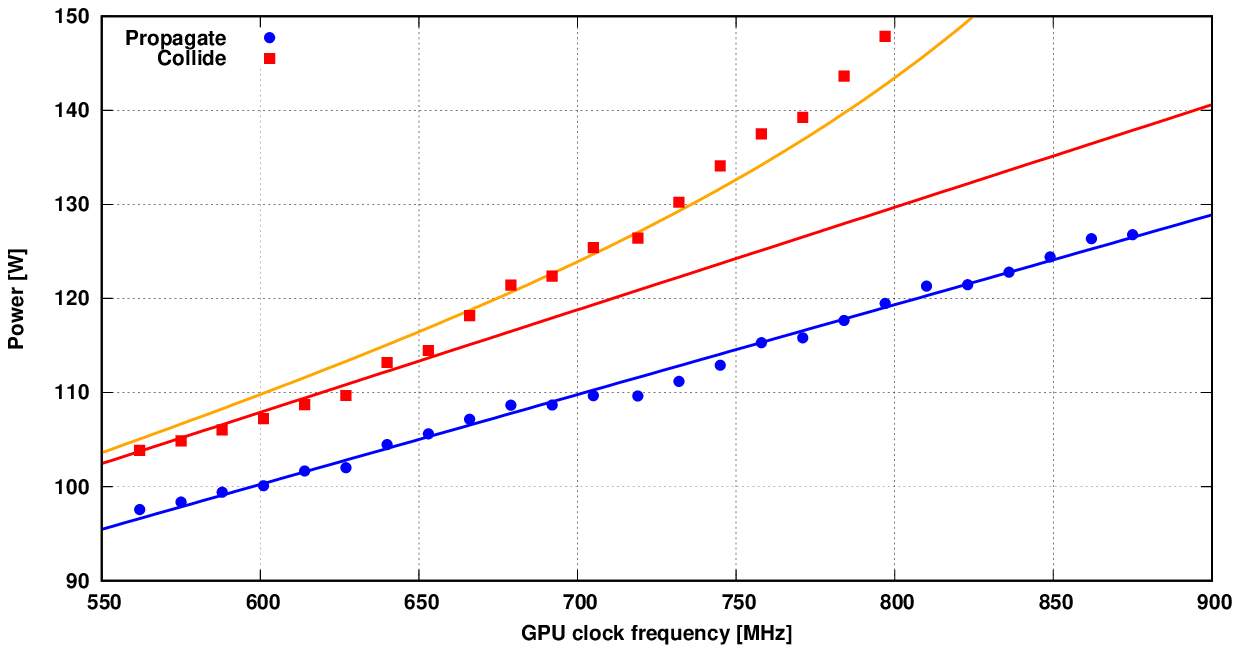}
\caption{Power measurements on a GPU, running the {\sf propagate} and
{\sf collide} kernels. the straight lines are the corresponding linear 
fits (Eqs. \ref{eq:propaEneFit} and \ref{eq:colliEneFit}), while the curve
is a fit of the non-linear power drain for {\sf collide} at $f \ge 650$~MHz 
(see text for details). }
\label{fig:eneFits}
\end{figure}

This analysis justifies the very good agreement of {\sf propagate} with our 
model: $E_S$ and $T_S$ decrease as we increase frequency and then (in our case 
at $f \ge 650$ MHz) increasing frequency further does not improve $T_S$ but 
increases $E_S$. The same behavior is qualitatively similar for {\sf 
collide} but not exactly as predicted. To further clarify the situation, we  
plot in Fig. \ref{fig:eneFits}  the power drained by the processor as a function of
$f$. For {\sf propagate}, we observe a linear 
behavior over the whole frequency range:
\begin{equation}
P (W) = 42.94 + 0.096~f.
\label{eq:propaEneFit}
\end{equation}
For {\sf collide}, the linear behavior is valid only for 
frequencies $f < 650$ Mhz. In this range a fit analogous to the previous one 
yields
\begin{equation}
P (W) = 42.50 + 0.109~f.
\label{eq:colliEneFit}
\end{equation}
As the clock frequency increases beyond $\simeq 650$ MHz, power drain for the 
{\sf collide} kernel grows much faster (we model the behavior in this range on 
purely phenomenological grounds as an additional contribution 
$P(f) (W)  = 0.005~e^{0.0099f}$); finally, the points on the plot at 
$f \simeq 800$ MHz and beyond do not describe correctly the actual situation, as
the clock  governor takes over frequency control, trying to keep the processor
within the allowed power budget; as a consequence the actual operating clock
frequency is most probably not the one requested by our code. 
This suggests that {\sf collide}  data points in Fig.~\ref{fig:timeVsFreq}
beyond $800$~MHz do not reflect a memory-bound regime.

Summing up, we conclude that our model describes accurately the
behavior of the {\sf propagate} kernel while for {\sf collide} we cannot be 
conclusive, as we enter a power regime at which we are unable to control the 
clock frequency before we enter the memory-bound regime.
This is related to the fact that {\sf collide} has a particularly high
\textit{operational intensity}: $I \approx 13.3$, while the two architectures 
have a theoretical maximum (i.e. computed taking into account maximum 
\textit{Turbo/Auto -boost} frequencies) $M_b$ of $5.61$ for the CPU and $8.08$ for the GPU.

\begin{figure}[ht!]
 \centering
\includegraphics[width=0.8\textwidth]{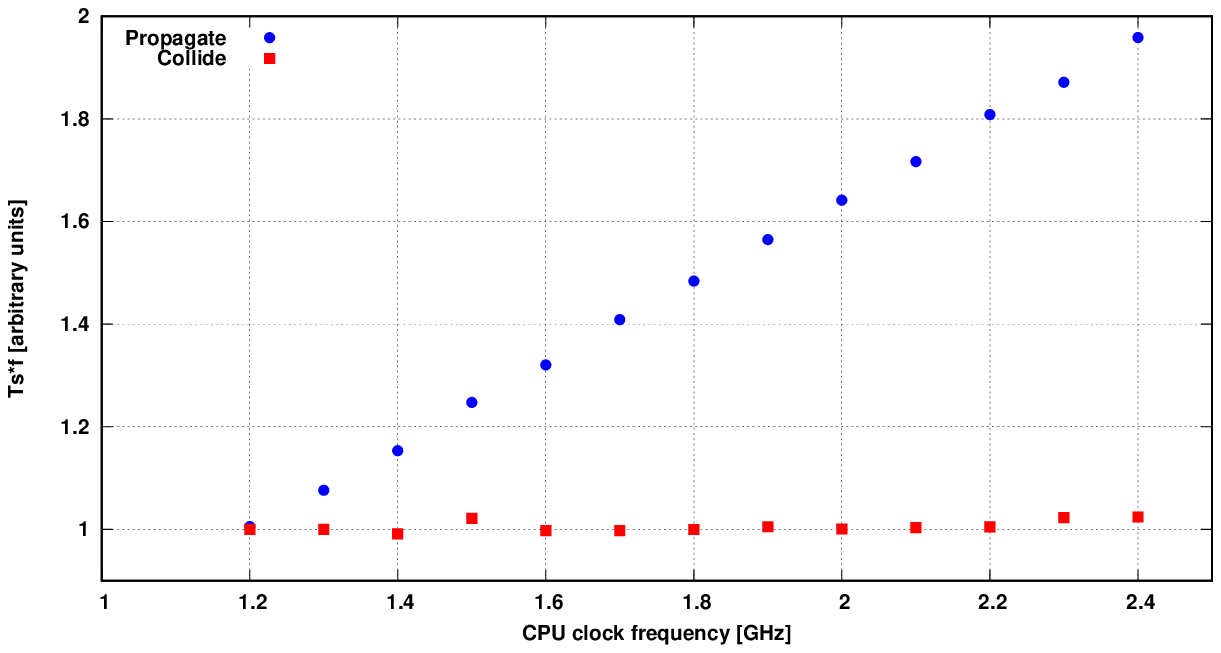}
\caption{
$(f~T_S)$ as a function of $f$ for {\sf propagate} and
{\sf collide}, running on a CPU, suggesting a memory-bound regime for 
{\sf propagate} and a compute-bound regime for {\sf collide} on the whole $f$ 
range.}
\label{fig:timeVsFreqCPU}
\end{figure}
Applying the same line of reasoning to our CPU data, our initial prediction is
that {\sf propagate} is memory-bound on the whole allowed frequency range 
(since $T_S$ is approximately constant) while {\sf collide} is not, in the 
frequency range that we control (since we see a constant correlated decrease of 
$E_S$ and $T_S$). We check that this is true in Fig. \ref{fig:timeVsFreqCPU}, 
showing again the behavior of $(f~T_S)$ as a function of $f$: we see that this 
quantity is linearly increasing for {\sf propagate} and approximately constant 
for {\sf collide}. Note that the best option to obtain the highest performance 
for {\sf collide} is to let the clock governor free to select the operating 
frequency; this presumably allow the system to enter \textit{Turbo Boost} mode, when
the power budget permits it, increasing $f$ up to 
$3.2$~GHz~\footnote{To manually set \textit{Turbo Boost} frequencies is not 
  permitted, so we could not perform in dept studies in this regime.}. 
This comes with some associated $E_S$ costs. 
However, the same strategy for {\sf propagate} has no advantage in 
performance while it still has some moderate $E_S$ cost. 
Finally we see that, for both routines, asking the clock governor to apply the 
\textit{powersave} strategy, enabling the namesake governor, has an adverse 
effect on both $T_S$ and $E_S$.

\section{Results and Discussion}
\label{sec:tradeoff}

Our results allow to draw some preliminary conclusions on possible hardware 
approaches to energy optimization; from Figs.~\ref{fig:gpu-et} 
and~\ref{fig:cpu-et} one can identify the performance-vs-energy tradeoff made
possible by varying the processor frequencies.
The first two lines of Table~\ref{tab:comparison} recap this 
information, listing the best energy saving made possible by these techniques
for the two analyzed functions, alongside with the corresponding performance
drop that one has to discount. 
Gains are not large but not negligible either; a simple lesson is that fair 
energy savings are possible by tuning the processor clock to
lower values in all cases in which the code is memory-bound.
Many HPC codes today are memory-bound, thus this simple recipe could be usable 
in a wide range of applications.

An aside result is a direct comparison between the two architectures: for 
example taking into account the best $E_S$ values, the 
CPU is $\approx 4.5$ times more energy demanding despite being $\approx 6.5$ 
times slower than the GPU to compute the {\sf collide} function.
This obviously refers only to this code, which is well suited for 
parallelization on GPU cores.

\subsection{Function by function tuning}

\begin{figure}[ht!]
 \centering
 \subfloat[\sf Propagate followed by Collide, CPU.]{
 \includegraphics[width=0.9\textwidth]{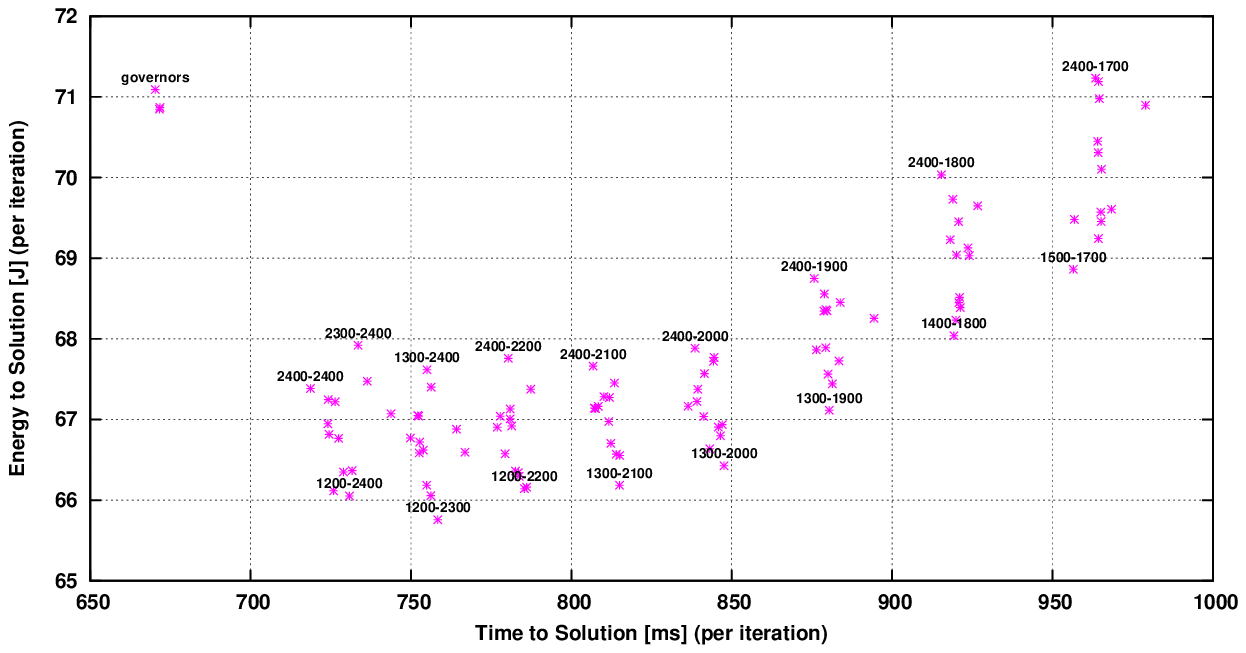}
   \label{fig:cpu-propcoll-et-zoom}
 }\\
 \subfloat[\sf Propagate followed by Collide, GPU.]{
 \includegraphics[width=0.9\textwidth]{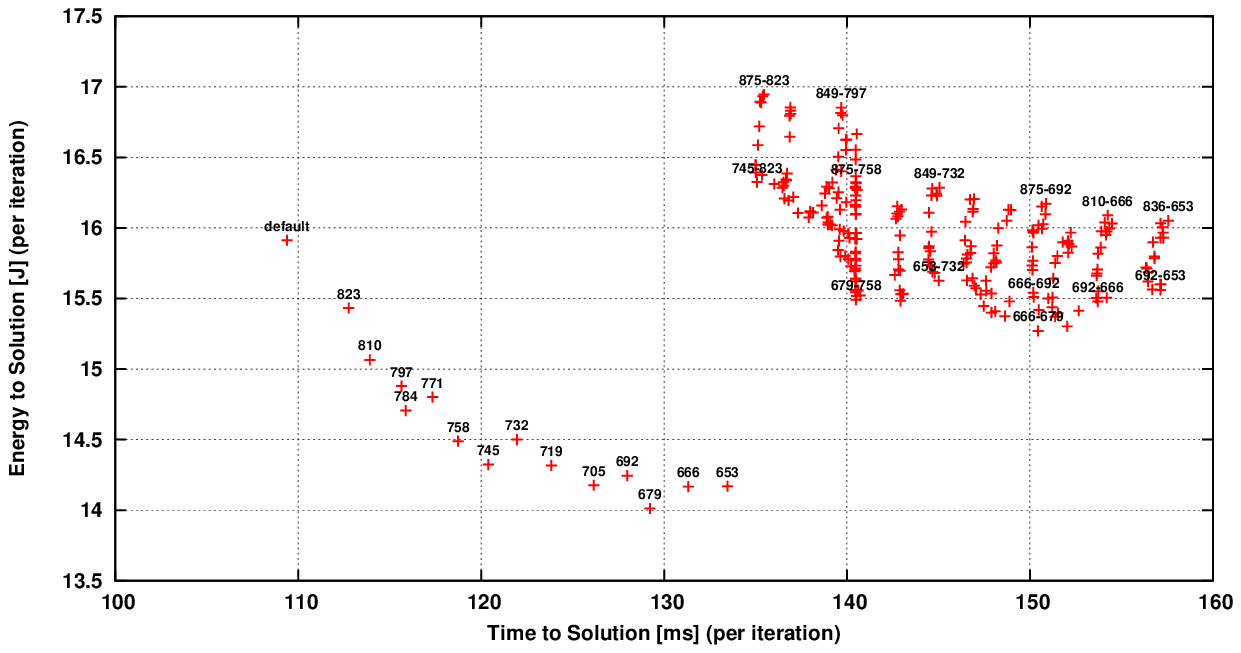}
   \label{fig:gpu-propcoll-et}
 }
\caption{$E_S$ vs $T_S$ for the {\sf propagate} function followed by the 
{\sf collide} function, changing the clock frequency on-the-fly.
Points are labeled with the used frequencies in MHz: left and right digits
represents respectively the frequency set while running the {\sf propagate} 
kernel and the {\sf collide} kernel.
A single frequency is reported if the same clock was used for both the 
functions.}
 \label{fig:propcoll-et}
\end{figure}

We now consider a further optimization step: one may try to 
select for each routine the optimal processor frequency, changing it on-the-fly
before entering each kernel; this approach may fail if 
tuning the clock introduces significant additional delays and/or energy costs.
Results are shown in Fig.~\ref{fig:propcoll-et}, 
plotting our measured values of $E_S$ vs $T_S$ for the same combination of the 
two routines that would apply in a production run, as the processor clocks are 
set to all possible values before entering each routine; for
CPUs this approach is indeed viable, as the overhead associated to on-the-fly 
clock adjustments is negligible. 
This is consistent with our independent 
measurement of the time cost of each clock change
($\approx 10 \mu s$) in agreement with similar measurements 
on Intel Xeon architectures~\cite{cpu-freq-lat}.
An interesting feature of the plot is that data points organize in vertical 
clusters, corresponding to sub-optimal clock frequencies for {\sf propagate} 
and clearly showing the adverse effects of a poor frequency match.

For GPUs we see immediately that this technique is not applicable with the 
available clock adjustment routines, as the time 
cost of each clock change ($\approx 10 ms$) is not negligible w.r.t. the average
iteration time.
We can anyway select a single clock frequency which is optimal (according to the
desired metric) for the whole code.

Table \ref{tab:comparison} summarizes costs and benefits comparing our best 
results (in terms of $E_S$ minimization) with the corresponding figures related 
to the default clock governors of the two processors; we see that limited but 
non-negligible improvements are possible for $E_S$ with a fair drop of the code 
performance.

\begin{table}[ht!]
\centering
\begin{tabular}{l|rr|rr}
\toprule
         & \multicolumn{2}{c}{GPU}   & \multicolumn{2}{c}{CPU} \\
Routine  & $E_S$ saving & $T_S$ cost & $E_S$ saving & $T_S$ cost \\
\midrule
{\sf propagate} & $18\%$  & $0$ & $9\%$ & $3\%$ \\
{\sf collide}   & $6\%$  & $10\%$ & $4\%$ & $4\%$ \\
\midrule
Full code   & $11\%$  & $10\%$ & $7\%$ & $8\%$ \\
\bottomrule
\end{tabular}
\caption{Approximate $E_S$ gains made possible by clock tuning and 
corresponding $T_S$ costs.\\
Percentages referred to $E_S$ optimal frequencies.}
\label{tab:comparison}
\end{table}

\subsection{Full code on multi-GPUs}

The adopted application, given its higher performance on these systems, is
commonly run on GPU based HPC clusters.
Therefore it may be interesting to measure how much the proposed 
energy-optimizations could impact on the energy consumption of a whole GPU-based 
compute node while running a real simulation.

With this aim, we run the original full code without any instrumentation on the
16 GPUs hosted on a single node of the COKA Cluster while measuring its power
consumption during the simulation execution.
The simulation was run on a $16384 \times 8192$ lattice for $10,000$ iterations.

Power measurements, in this case, were performed using the IPMI protocol, in
order to acquire data directly from the sensors embedded in the power supplies
of the Supermicro SYS-4028GR-TR system.
The full node maximum declared power consumption is $3.2$kW. 
It hosts 8 NVIDIA K80 board accounting for a maximum of $300$W each and thus 
$2.4$kW in total, giving a $75\%$ of the maximum power drained possibly only 
by GPUs. 

Given that we demonstrated that clock tuning on a function-by-function basis
is not convenient for GPUs at this stage, we run the full simulation at 
different fixed GPU clock frequencies, including one run with the default 
frequency governor with \textit{Autoboost} enabled.

\begin{figure}[ht!]
 \centering
 \subfloat[\sf Full node power drain, setting all the GPUs at the same clock 
 frequency, for different clock frequencies. Power drain measured by the power 
 supplies through IPMI.]{
   \includegraphics[width=0.9\textwidth]{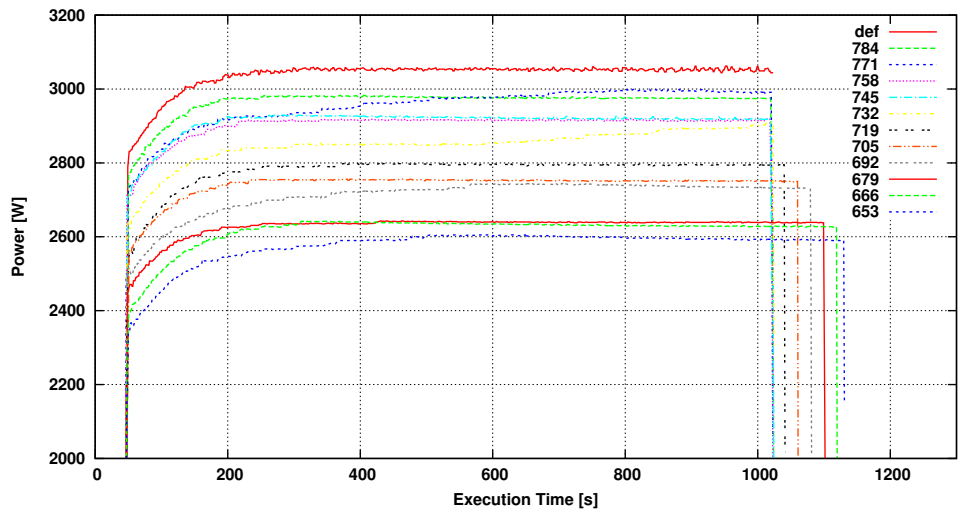}
   \label{fig:node-power}
 }\\
  \subfloat[\sf Energy derived from average power drained and execution time,
  normalized per iteration.  
  GPU clock frequencies as labels in MHz.]{
   \includegraphics[width=0.9\textwidth]{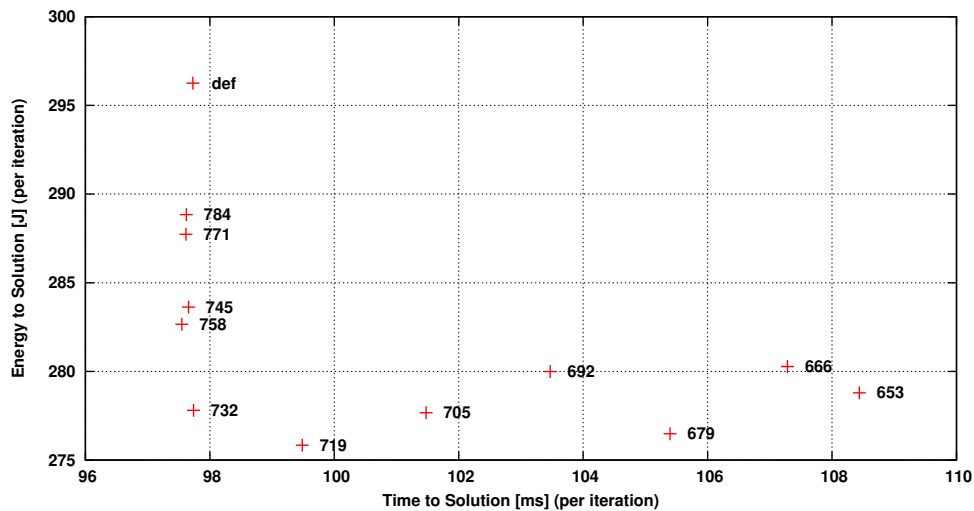}
   \label{fig:node-energy}
 }
 \caption{Running 10k iterations of a complete Lattice Boltzmann simulation over
  a $16384 \times 8192$ lattice, on a single cluster node, using 16 GPUs, for 
  different GPU frequencies.}
  \label{fig:node}
\end{figure}

Power readings for the full node are reported in Fig.~\ref{fig:node-power}, 
while the corresponding $E_S$ for the application run are reported in
Fig.~\ref{fig:node-energy}.

As can be seen from Fig.~\ref{fig:node-energy} at a specific frequency (i.e. 
732MHz) $\approx 7\%$ of the total consumed energy of the computing
node can be saved, with respect to the default behavior, without impacting 
performances.


\section{Conclusions}
\label{sec:conc}

In this paper we have analyzed the potential for energy optimization of an 
accurate selection of the processor clock, tuning
\textit{machine balance} of the processor to \textit{operational 
intensity} of the application.
We applied this method to a typical HPC workload on state-of-the-art 
architectures, discussing the corresponding possible performance costs. 

Our approach yields sizeable energy savings for typical HPC workloads with a 
simple and fast tuning effort; the associated performance costs are easily 
manageable. 
The main advantage of this approach is that it can be applied to any production
program with a limited amount of easily-applied measurements, not involving 
additional hardware, or expensive program adjustments.

Some further remarks are in order:
\begin{itemize}
\item It is relatively easy to instrument application codes to have accurate
energy measurements of key application kernels, using hardware counters and 
minimally disrupting the behavior and performance of the 
original code.
\item Simple theoretical models -- while admittedly unable to capture all 
details of the $E_S$ and $T_S$ behavior --  still provide a level of 
understanding that is sufficient to guide optimization 
strategies.
\item An important result is that the best practice for both performance and
energy efficiency is to look at the $I$ and $M_b$ parameters: if $I
\leqslant M_b$, reducing $M_b$ (e.g. lowering the clock frequency) will reduce
$E_S$ with a negligible impact on performance;
in other words, for compute-bound codes the best strategy is to run the code at 
the highest sustained processor frequency
while, for memory-bound code, decreasing the clock frequency usually
reduces $E_S$ with essentially no impact on performance. 
Since many real-life HPC codes today are memory-bound, this practice 
could be useful for many applications. 
\item Default frequency governors usually yield the best performances, but
do not minimize energy costs; on the other hand, using the lowest 
frequencies (such as prescribed by the \textit{powersave} cpufreq governor), 
is in general a bad choice, from the point of view of both $E_S$ and $T_S$. 
\item ``on-the-fly'' clock adjustment is a viable strategy 
on recent Xeon processors, while on NVIDIA GPUs the overheads of 
the clock management libraries makes this strategy hardly useful.
\item Highest frequencies, in particular the ones enabled by \textit{Auto/Turbo -boost} 
mechanisms, let power drain increase superlinearly with respect to $f$ causing 
a systematic divergence from our predictions.
This may be due to higher current values within the processor at higher 
operating temperatures, making modeling and predictions less solid, but also 
puts a question mark on operating HPC machines at high temperatures in an 
attempt to reduce cooling costs.
\end{itemize}

We plan additional experiments in this direction for multi-node implementations 
in order to estimate the energy-saving potentials at the cluster level, taking 
into account also communications costs.
We also plan to develop more detailed power-performance-energy model able to
predict more precisely the behavior of these variables on modern architectures.

\subsubsection*{Acknowledgements}
This work was done in the framework of the COKA, COSA and Suma projects of INFN.
We would like to thank all developers of the PAPI library (and especially V. M. 
Weaver) for the support given through their mailing-list.
All benchmarks were run on the COKA Cluster, operated by
Universit\`a degli Studi di Ferrara and INFN Ferrara.
A. G. has been supported by the European Union Horizon 2020 Research and 
Innovation Programme under the Marie Sklodowska-Curie grant agreement 
No.~642069. 


\bibliographystyle{wileyj}

\end{document}